\documentclass[english, 12 pt]{article}
\usepackage[T1]{fontenc}
\usepackage[latin9]{inputenc}
\usepackage{graphicx}
\usepackage{babel}
\usepackage{url}
\usepackage{multicol}
\usepackage{amsmath}
\usepackage{rotating}           % for sideways tables/figures
\usepackage{color}

\makeatletter

\newcommand{\fig}[1]{Figure(\ref{#1})}

\newcommand{\la}[1]{ \label{#1}}

\newcommand{\e}{\epsilon}

\newcommand{\bsubs}{\begin{subequations}}
\newcommand{\esubs}{\end{subequations}}

\providecommand{\ba}{\begin{align}}
\providecommand{\ea}{\end{align}}

\usepackage{multirow}
\newcommand{\be}{\begin{equation}}
\newcommand{\ee}{\end{equation}}

\newcommand{\bea}{\begin{eqnarray}}
\newcommand{\eea}{\end{eqnarray}}
\providecommand{\work}[1] {\color{black} #1}

\begin{document}

\title{ Reflections on Gibbs: \\From Statistical Physics to the Amistad  V3.0 \\
}

 \author{ Leo P. Kadanoff\footnote{e-mail:  lkadanoff@gmail.com}~
 \\
\\
 The James Franck Institute\\
The University of Chicago
\\ Chicago, IL USA 60637
\\ 
\\}
\date{date: originally drafted and delivered February, 2003,\\ revised March, 2014} 

\maketitle

\begin{abstract}
This note is based upon a talk given at an APS meeting in celebration of the achievements of J. Willard Gibbs. 
 
J. Willard Gibbs, the younger, was the first American physical sciences theorist. He was one of the inventors of statistical physics. He introduced and developed the concepts of phase space, phase transitions, and thermodynamic surfaces in a remarkably correct and elegant manner. These three concepts form the basis of different areas of physics. The connection among these areas has been a subject of deep reflection from Gibbs' time to our own. This talk therefore  celebrated Gibbs by describing modern ideas about how different parts of physics fit together.

I finished with a more personal note. Our own J. Willard Gibbs had all his many achievements concentrated in science. His father, also J. Willard Gibbs, also a Professor at Yale, had one great non-academic achievement that remains unmatched in our day. I  describe it.
\end{abstract}

\newpage

\tableofcontents
\newpage
\renewcommand{\theequation}{1-\arabic{equation}}
\setcounter{equation}{0}
\section{How do theories fit together?}
Older ideas suggest that physical theories merge via a limiting process in which some parameter smoothly approaches its final value. Thus Newtonian mechanics
might emerge as a limit of special relativity's in the limit as the latter is applied to objects moving at low speeds.  This view seems acceptable for the connection between classical mechanics  and special relativity, but appears to be just a bit too tame, and
not quite right, for such examples as the relation between classical mechanics and quantum theory\cite{Bokulich,CRV} or the relation between special relativity and general relativity.   A newer view, which I describe as a theory of  ``singular connection,''  was put forward by the work of Michael Berry\cite{Berry4,Berry}  and then expressed inphilosophical terms by Robert Batterman\cite{Batterman, Bokulich}. This view replaces\cite{CRV} the metaphor of ordinary limits\cite{Nagel} by one based upon the applied mathematics ideas of asymptotic behavior and of singular perturbations.
\subsection{Singular Connections}
The relation between quantum theory and classical mechanics is a amazingly rich and varied subject that has been analyzed in detail by Michael Berry\cite{BerryQM, Berry2} and Alisa Bokulich\cite{Bokulich}.  At least three different conceptualizations, based upon different ideas, describe highly quantum behavior, ordinary classical mechanics and the ``classical limit'' region in between.  However, instead of exploring this intricate subject, I shall introduce the singular connections approach using examples close to Gibbs: the Gibbs phenomenon in Fourier series and the relations among the parts of statistical physics.

\section{J. Willard Gibbs' scientific contributions}
He was a theorist interested in the mathematical and theoretical structure of his problems. He was extremely careful to publish and utter only true things about physics, chemistry, and mathematics.  His  scientific contributions include: 

\begin {itemize}
\item{Thermodynamics: The thermodynamic limit.}  Gibbs produced a very careful and full description of thermodynamics\cite{GibbsThermo}.  He  particularly focused upon the description of how thermodynamics arises in the limit of large homogeneous systems, when they are left undisturbed for a sufficient time for them to relax to equilibrium.  This limiting situation is known as the {\em thermodynamic limit.}  
\item{More thermodynamics:} Gibbs also invented the {\em phase rule} which described the equilibrium among phases. Still again, he invented thermodynamic surfaces and chemical potentials. He recognized that thermodynamics is true independent of the microscopic models that might be used to justify it.
\item{Statistical mechanics:} Along with Ludwig Boltzmann,  Gibbs invented the use of phase space and ensembles as the basis of statistical mechanics\cite{GibbsStat}.  
\item{Vector calculus:} Gibbs invented and  introduced dot and cross products.  This looks like a minor contribution now, but it was quite important at the time.
\item{Understanding vectors:} Apparently Gibbs held to the modern idea that a vector was a thing in itself, while his contemporaries mostly confused the components with the object.
\item{Gibbs overshoot:}  In this talk,  I shall discuss in some detail the  ``Gibbs Phenomenon'' in Fourier theory\cite{Gibbs1,Gibbs2}. (This behavior was earlier discovered by Wilbraham\cite{Wilbraham}, but Gibbs made it generally known.)
\end{itemize}

\subsection {Gibbs and the missing connection.}
 He thought particularly clearly and deeply about phase transitions in the context of thermodynamics, but he hardly mentions them in his later book on statistical mechanics.  Arthur Wightman said\cite{Caldi}[page 34]
\begin{quotation}
There is one aspect of the thermodynamic limit that Gibbs does not emphasize. That is the appearance of phase transitions between distinct thermodynamic phases. [S]harp phase transitions do not occur in finite systems. A little more pedagogical zeal by Gibbs could have saved some of the generations that followed considerable time. [... ] Gibbs could have told them that it was pointless to discuss [phase transitions] except in the thermodynamic limit.
\end{quotation} 
It is, in some sense, very surprising that Gibbs did not emphasize that  statistical mechanics and thermodynamics came together in the description of phase transitions. These transition provide an excellent example for understanding the connection between these two areas of science which he helped invent. 

The connections among various parts of science is a very deep subject that has been of interest to both philosophers and physicists.  The technical word used by both physicists and philosophers for this process of connection is {\em reduction}. Unfortunately they use the word differently.  One group would insist that quantum theory reduces to classical while the other would suggest that reductions works in just the opposite sense. 	I'll stick with the word {\em connection}  which is more neutral.  

\section{Connections}

\subsection{Simple limiting process}
Sometimes the connections are very simple. The pendulum's motion becomes identical to that of a simple harmonic oscillator whenever the pendulum oscillates weakly and remains close to its equilibrium position. The connection involves a low-energy limit of the pendulum and a limiting process just about as simple as the one by which the calculus' derivative operation is defined. The concept of a limit is not fully transparent to an unprepared mind. (Think of the troubles of Bishop Berkeley\cite{Analyst} with derivatives, or of many generations with the limiting processes involved in Achilles catching up with and then passing the tortoise.) However, it is now familiar and direct, so that all theory connections which involve only the simple version of the concept of limits may be said to be well understood.

\subsection{Something missing/extra}
But the concept of a simple limit does not seem to describe many of the limiting process that occur in relating different domains of physical science. In doing a simple limit, you don't get anything really new, the old thing morphs smoothly into its limiting form. However think of our usual examples of theory connection as exemplified by the listing in Table (1).)

Some examples: 
\begin{itemize}
\item  The richest case is the first one in the table, outlining the amazing connection between classical mechanics and quantum mechanics\cite{Berry2}.  These two approaches conceptually quite different: classical mechanics describes the motion of particles through trajectories, quantum theory uses wave functions and probabilities.    Very roughly speaking, the size of Planck's constant, $\hbar$ controls the relative importance of quantum effects.   The intermediate case, the quasi-classical domain contains a host of phenomena not belonging to the two extremes.  Tunneling and the chaotic spacing of higher nuclear energy levels  are just two of many known quasi-classical effects. 
\item   In the next example, in the second row, physical optics\cite{Berry2} works when the observation tools are quite gross in comparison to the wavelength of light. Then, light behaves as a set of rays with well-defined, particle-like, trajectories.  On the other hand, at the finest level, we understand that light is electromagnetic radiation and solve wave equations to understand its behavior. In the intermediate case,  we have all the effects of fringes, rainbows,  and diffraction.  One would not have guessed this, often fractal, complexity from a brief examination of either the wave equation or ray trajectories\cite{Berry2}.  
\item  In another example, Newtonian mechanics contains a concept of absolute time. It applies to objects of everyday size.  General relativity describes the effects of gravity at the very largest scales, those that apply to the cosmos.  In general relativity, time is just a part of a complex geometrical structure, often with an interesting topology.  Between these limiting cases, we might find a ray theory of gravitational waves, not Newtonian, but also very far from the considerations of non-Euclidean geometry put together by Einstein.  In addition, we can expect gravitational waves to exhibit, at the very least,  all the complexity shown by wave-interference in optical situations.  
\end{itemize}  
In short, the limiting theories do not talk the same language, and neither talk the language of the intermediate case.

\subsection{Connections}
% chaos and determinacy coexist bifurcations catastrophes
%Connections
All authors accept  that connections among theories are done with limiting process, like taking $\e$ to zero. However Batterman and Berry argue that the right ideas of limits are borrowed from the concepts of singular limits and asymptotics of Twentieth Century  mathematics\cite{Hinch}, not the simple limits of Newton and Liebnitz.   Part of their view:

{\bf Concept I: singular perturbation: sometimes even a little $\e$ is a lot.} Example: 
The equation, $x-1 = 0$ has but one solution, $ x=1$,  but an allied  perturbation problem can have two or more solutions
For example, 
$$x-1 = \e~x^2$$
\begin{itemize}
\item first solution, for small $\e$,  $x=1+\e +2\e^2 + ...$ 
\item second solution, for small $\e$,  $x=\e^{-1}-1 -\e+..$.
\end{itemize}
Here the small parameter controls the number of solutions. (A singular perturbation may be expected whenever the highest order term in an equation is multiplied by a small parameter.)
Typically the interesting connections  among physical theories arise from singular perturbations.    For example, $\hbar$ multiplies the highest order derivative term, 
$ \hbar^2 \nabla^2/(2m)$,  in the Schr\"odinger equation.   For light waves, one can see the small-parameter behavior by manipulating the wave operator to read ${\partial_t}^2 - \nabla^2/c^2 $ and then taking  $1/c$ to be the small parameter.   Then, it too multiplies the highest order space derivative.

\begin{table}
\begin{tabular}{|c| c| c |}
\hline
first theory  &  second theory& intermediate case \\
\bf{typical concept} &\bf{typical concept}  & \bf{typical concept}  \\
\hline
\hline
classical mechanics & quantum mechanics & ``classical limit''  \\
\bf trajectories & \bf entanglement &\bf  tunneling \\
\hline
wave optics & physical optics & limiting behavior  \\
\bf waves & \bf rays &\bf fractal patterns\\
\hline  
thermodynamics & statistical mechanics & critical behavior \\
thermodynamic limit & finite $N$  & infinite system  \\
\bf discontinuous jumps & \bf smooth behavior &\bf  algebraic singularities\\
\hline  
Newtonian mechanics & general relativity & intermediate case \\
\bf absolute time & \bf interesting geometry &\bf  graviton optics\\
\hline
molecular beam & fluid dynamics & molecular gas \\
\bf orbits & \bf hydrodynamics &\bf  mean free path\\
\hline
\end{tabular}
\centering 
 \label{connection}
\caption{Connection among different theories.  Different connections are described by different rows, with the two limiting cases first, and the intermediate case given last.   That situation turns out to be quite different from the limiting cases and needs conceptually different tools for its analysis.  These cases have  been elucidated by centuries of scientific work.      } 
\end{table}

{\bf Concept II. The asymptotic expansion.} Usual power series like the familiar one
$$
e^x =1 + x +x^2/2! +...+ x^n/n! + ....
$$
define an interesting but  smooth behavior.  You use the expansion by computing successive terms and adding them up. The more terms you add, the better is your answer. There is nothing left over, hard work will give you the whole function.
Consider, in contrast, a typical asymptotic expansion like the Stirling approximation for $n!$,
$$
\ln (n-1)! = n \ln n -n +0.5 \ln n - \ln (2 \pi )^{1/2} +O(1/n) +...
$$
For all $n$ you get a pretty good answer from the first term. For higher $n$, you get a better answer by including more terms. But for each n, you should stop somewhere: there is an optimal number of terms to compute. If you go further you get a worse answer.   An asymptotic analysis is accurate, as far as it goes, but can leave out something. For
example, in the expansion of $\ln (n^2+ n!)$, we would
never see the $n^2$ in the asymptotic analysis because there are an infinity of larger terms.

\subsection{A story}
The late Tony Houghton and I were doing a renormalization group calculation very early in the history of such calculations. We had developed a theory and we were calculating a critical index called $\nu$ which described how the correlation length diverged as the temperature approached the critical temperature in the two-dimensional Ising model.
Our zeroth order theory gave {\work $\nu_0=0.7$}. The exact answer, according to the Onsager exact calculation
was $\nu_\text{exact}=1$. So our zeroth order answer was not too awful! After some work we got a first order result, {\work $\nu_1=0.9$}.	Not bad!  More work gave {\work $\nu_2=0.99$}, better! Then, after much work we got the third order {\work $\nu_3=0.999$}. Great!   We saw that we
could, by working very hard, just accomplish the next order. We did it and finished the calculation, by finding {\work $\nu_4=1.7 $} !....... We had bumped into an asymptotic
expansion! Some physics was left out, and that left--out physics prevented us from ever getting to the exact right answer.  In fact, if we had calculated more terms it is likely that our answer would simply have continued to get worse.  

So ?

Typically singular perturbations describe situations in which there are small  parameters, and quite different behaviors in different regions of the parameters\cite{Hinch}. They can often be effectively analyzed by using different asymptotic expansion in different regions of the behavior. Each expansion catches a part of the truth for that problem and leaves out another part. The expansions are conceptually different and each one leaves out an interesting part of the physics. 
In contrast, ordinary power series expansions describe smooth behavior and contain all the physics in one go.  As you include more terms you get more accurate answers, but nothing really new happens since no qualitative error was made at any point. 
 
The right metaphor for connecting phase transition to the statistical mechanics of finite systems is not the power series expansion and its smooth approach to a limit.	According to Berry and Batterman and much contemporary thought it is the asymptotic expansion which is demanded in the analysis of singular perturbations.
Berry\cite{Berry} showed the asymptotic analysis gives an interesting and worthwhile view   of waves,   quantum chaos, and many other situations. Batterman\cite{Batterman} developed a philosophical discourse on this subject. Let me give a little analysis of a simple problem, due to J. Willard Gibbs, the younger.
\section{Gibbs' analysis; Gibbs' error}

Consider the Fourier series
$$
f_N(x) = -4 \sum_{n=1}^N \frac{(-1)^n}{n} \sin(nx)   
$$
in which the series is cut off for large but not infinite values of $N$. Think of the analysis of this function as a simple example of a physics theory. Since the function in question is the Fourier series expansion of the periodically repeated function
$$
f(x)= \begin{cases}\cdots,\\ 2x & \text{ for }  -\pi<x<\pi   \\ 2 x-4\pi & \text{ for~~}  \pi<x<3\pi  
 \\ 2 x-8\pi & \text{ for~~}  3\pi<x<5\pi
 \\
 \cdots. \end{cases}
$$
The function being transformed looks like a sawtooth  as in \fig{sawtooth}. Note that this sawtooth has $2 \pi$ as its maximum possible value, which value is achieved at the end of each interval.  

 In the limiting case, $N\rightarrow \infty$, we might wish and expect to get different  "theories", for each interval,  \dots,  $(-3\pi,-\pi),(-\pi,\pi), (\pi,3\pi),$ \dots
In his first work on this subject\cite{Gibbs1}, Gibbs studied the curves $y=f_N(x)$ for the different
values of $N$. He then said, in a {\em Nature} paper\cite{Gibbs1}, that the curves approach the curve $y=f(x)$ as $N$ goes to infinity. This is an error, as Gibbs soon recognized, and published as an erratum\cite{Gibbs2} in {\em Nature}. 

%%%%
\begin{figure}[!] %\begin{centering}
%\begin{multicols}{2}
%\begin{center}
\includegraphics[width=14 cm ]{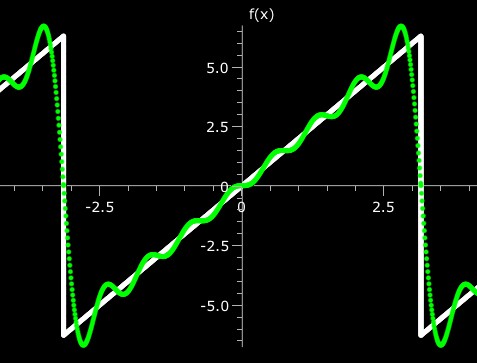}
%\end{center} 
\caption{The function $f(x)=f_\infty(x)$ is a sawtooth repeating itself with an $x$ interval of $4 \pi$.  If we write down a Fourier series constructed from this function, it follows the general path of the sawtooth, but has wiggles in it. Here we show a series with eight terms.   The question for Gibbs is ``how does the series approach the sawtooth for large $N$''. Here we see near the extrema of the sawtooth an overshoot of the approximate function which goes a bit above and below than the extrema.   This  overshoot does not diminish for  larger $N$.  }
 \la{sawtooth}
\end{figure}

A more careful analysis of the large $N$ behavior is necessary.   An asymptotic analysis of this situation could be described by different ``theories'' of the series, $f_N(x)=x$, in different regions of $x$, i.e..
\begin{itemize}
\item {\bf first ``theory:''}  For $x$ is in the interval $(-\pi,\pi)$ except very close to the ends, $f_N(x)$ is
equal to $2 x$.  (See \fig{Unscaled}.)
\item   {\bf  second ``theory:''} For $x$ is in the interval $(\pi,3\pi)$ except very close to the ends, $f_N(x)$ is
equal to $2(x-2\pi)$.
\item  {\bf  third ``theory:''}   Very close to the boundaries a different kind of theory is required; one which takes into account the wiggles in $f_N$ that occur in this region. As $N$ get larger the maximum height of the wiggles does not decay but instead approaches a fixed value. Note that, for larger $N$, the wiggles get more closely bunched at the endpoints of the interval. (See \fig{Unscaled}.).   To see a workable theory of this effect plot $f_N$ as a function of a variable that gives more sensitivity as one goes to larger $N$.  For example for the endpoint at $x=\pi$ one could use the variable $z=\pi + (x-\pi)N$, as in \fig{Scale}. Then, as $N$ get larger, $f_N(z)$ approaches a constant form, called a {\em scaling function}.  Indeed this scaling function does get larger than $2 \pi$.  This  analysis provides a third ``theory'' of the overshoot, conceptually close to modern theories of critical phenomena, scaling, and universality\cite{Widom,Kadanoff,Domb}.   
\end{itemize}
%Note that the different curves  all overshoot $\pi$. This overshoot approaches a limit that larger than $\pi$ for large $N$.  Since $\pi$ is  the value that Gibbs claimed to be the limit of the series at the end point, Gibbs' analysis is clearly wrong.    
The overshoot behavior just described has become known as a {\em Gibbs overshoot} or a {\em Gibbs phenomenon}. The fact that asymptotic analysis gives different behaviors in different regions was first presented by George Stokes\cite{Stokes}.
%\subsection{Gibbs'  analysis}

\begin{figure}[h] %\begin{centering}
%\begin{multicols}{2}
%\begin{center}
\includegraphics[width=11 cm ]{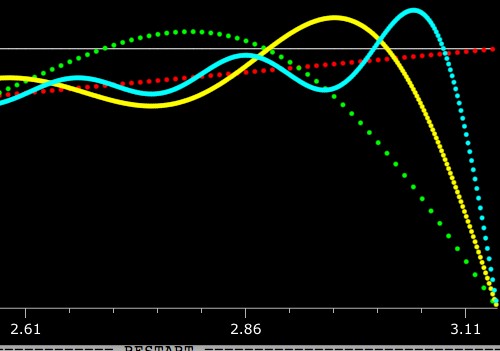}
%\end{center} 
\caption{For large $N$, the function $f_N(x)$ does not look like a sawtooth in the region  near the ends of the sawtooth's intervals of monotonic rise.  Here we have a blown-up view of this region with plots of $f_N(x)$ for various different values of $N$. The range of  $x$ in this view  is from $x=0$ to approximately $x=7$.  The thin white line is the incorrect ``Gibbs limit'' as $x=2 \pi= 6.28 \dots$. The various curves are respectively:  in red we have the sawtooth at $N=\infty$, in green $N=8$ is plotted, in yellow $N=16$, and in cyan $N=32$.  Notice that as $N$ increases we see more closely spaced oscillation and a slow increase in the magnitude of the overshoot.  The overshoot reaches a maximum value of 0.562\dots at $N=\infty$. }
 \la{Unscaled}
\end{figure}

\begin{figure}[!] %\begin{centering}
%\begin{multicols}{2}
%\begin{center}
\includegraphics[width=14 cm ]{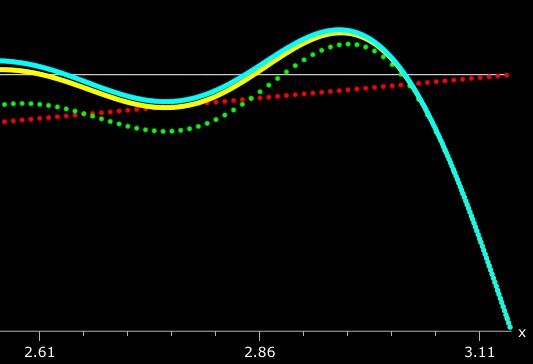}
%\end{center} 
\caption{The function $f_N(x)$ is plotted against $ z=\pi+ (x-\pi) N $ to give  a picture of the behavior near the ends of the interval.  This variable gives  a higher magnification in the argument of the function
 for higher values of N. Once again, the thin white curve is at $y=2\pi$ while the range of the entire y-axis is from zero to about seven. Note that as $N$ is increased, the maximum height approaches a constant somewhat larger than $2 \pi$.  The various curves are respectively:  in red we have the sawtooth plotted as a function of $x$, in green $N=8$ is plotted, in yellow $N=32$, and in cyan $N=128$, all three plotted against $z$.  As $N$ becomes larger the wiggles become more closely spaced,  and toward the endpoints of the intervals, more pronounced.  For the largest $N$, the curves in a plot like this would approach a constant form, slightly higher than the cyan curve.  This limiting result is called a {\em scaling function}.
}
 \la{Scale}
\end{figure}

\subsection{ Conclusion drawn from overshoot analysis}

From Gibbs' own example: A ``physical theory'' of the  process of convergence of discontinuous Fourier series requires different behaviors and different conceptualizations in different regions. This is because the $N$ goes to infinity limit produces, in essence, a singular perturbation and requires an asymptotic analysis.
In a much more sophisticated sense, this same story describes the connection between wave optics and ray optics, between classical mechanics and quantum theory,   and among a wide class of other problems as shown in the analyses of Berry\cite{Berry} and Batterman\cite{Batterman}.
In a slightly different way, the modern theory of fluctuation-dominated phase transition behavior is an intermediate asymptotics between Gibbs' theory of thermodynamic phases for infinite systems and Gibbs' probablisitic and ensemble related theory of statistical equilibrium in finite dimensional phase space. The phase transition theory contains ideas of scaling and universality simply absent in thermodynamics and the usual statistical mechanics.  The required scaling analysis, originally dues to Widom\cite{Widom},  has formed the basis for the modern theory of critical phenomena\cite{Domb}.  One might infer from Wightman\cite{Caldi} that Gibbs was perhaps close to seeing phase transitions and critical phenomena in the way we do today.   From the overshoot example, I would conclude, on the contrary, that although he was very careful not to make the mathematical mistake of thinking that a phase transition could occur in a finite system, Gibbs was not expecting the richness that arises in the region intermediate between different theories.  For that reason, he and his times were quite far from seeing how finiteness would affect behavior near phase transitions.

\section{Prehistory: The Amistad:}
Even more than most academics, J. Willard Gibbs, Jr. led an uneventful life, far from the great events of his day\cite{Wheeler}. (Of course, as Muriel Rukeyser\cite{Rukeyser} points out, his ideas were in themselves a great event of his era, but that's different.)  His father J. Willard Gibbs, Sr. also a Yale academic, a philologist, also spent almost all of his life far from the press of real life.  However, the senior Gibbs had one notable real-world achievement, connected with the arrival of the rebellion-torn slave ship Amistad from Cuba\cite{Jones}. President Van Buren wanted the mutineers to be considered, in modern terms, terrorists and sent back-- without trial-- to Cuba. Since nobody seemed to know their language, a return without a real hearing appeared inevitable. But Gibbs used gestures to find out how they counted and called out the number-words thus garnered at the port of New York. In this way, he found two translators. The translators gave voice to these captives and enabled them to convince the courts of their innocence.	Thus, the senior J. Willard Gibbs gave a trial and justice to these accused terrorists. Nobody, to this day, has equalled this achievement for our prisoners now being held in Cuba.

\section*{Credits}
I have had useful discussions about this work with Michael Fisher, Steve Berry, Marko Kleine Berkenbusch, Robert Batterman, and Michael Berry. I am indebted to Amy Kolan and her senior class at St. Olaf for helpful critical comments.
Research supported in part by NSF-DMR and also by the University of Chicago MRSEC and NSF grant number  DMR-0820054.

\end{document}